\documentclass[12pt,a4paper]{article}
\pdfoutput=1
\usepackage{jheppub}
\usepackage{tikz}
\usepackage{subcaption}
\usepackage{physics}
\usepackage{braket}
\usepackage[normalem]{ulem}

\usepackage{multirow,graphicx,amssymb,url,mathrsfs,amsmath}
\usepackage{wrapfig,boxedminipage,epsfig}
\usepackage{amsxtra,amstext,latexsym,dsfont,amsfonts}
\usepackage{color}

\begin{document}

\begin{flushright}
\end{flushright}
\title{A note on the non-planar corrections for the Page curve in the PSSY model via the IOP matrix model correspondence}

 \author[a, b]{Norihiro Iizuka}
 \author[c]{and Mitsuhiro Nishida}
\affiliation[a]{\it Department of Physics, National Tsing Hua University, Hsinchu 30013, Taiwan}
\affiliation[b]{\it Yukawa Institute for Theoretical Physics, Kyoto University, Kyoto 606-8502, Japan}
\affiliation[c]{\it Department of Physics and Photon Science, Gwangju Institute of Science and Technology, \\
Gwangju 61005, Korea}

\emailAdd{iizuka@phys.nthu.edu.tw}
\emailAdd{mnishida124@gmail.com}

\abstract{We develop a correspondence between the PSSY model and the IOP matrix model by comparing their Schwinger-Dyson equations, Feynman diagrams, and parameters. Applying this correspondence, we resum specific non-planar diagrams involving crossing in the PSSY model by using a non-planar analysis of a two-point function in the IOP matrix model. We also compare them with Page's formula on entanglement entropy and discuss the contributions of extra-handle-in-bulk diagrams.}

\maketitle

\section{Introduction}
Understanding the evaporation process of black holes \cite{Hawking:1975vcx} has played an important role in our understanding of quantum gravity. Since quantum gravity is a challenging research subject, the approach of studying toy models of quantum gravity is beneficial. 
It is essential to investigate the quantum effects in such toy models where we can compute quantum correction exactly. The Penington-Shenker-Stanford-Yang (PSSY) model \cite{Penington:2019kki}, sometimes called the West Coast model, is a very nice toy model of evaporating black holes. It is a 2-dimensional Jackiw-Teitelboim (JT) gravity, with end-of-the-world (EOW) branes. 
This model has two subsystems, a black hole with its Hilbert space dimension $e^\textbf{S}$ represented by the flavors of the EOW branes, and an auxiliary reference system $\bold{R}$ representing ``radiation'' with its Hilbert space dimension $k$. It was shown in \cite{Penington:2019kki} that depending on $k < e^\textbf{S}$ which corresponds to early black hole or $k > e^\textbf{S}$ which corresponds to late black hole, the dominant topology in the gravitational path integral changes and that leads to the Page curve behavior changes before and after the Page point \cite{Page:1993df, Page:1993wv}. 

The analysis done by \cite{Penington:2019kki} above is in the planar limit and to see the essential Page curve behavior change at the Page point, this planar analysis is good enough. However, it is certainly interesting to investigate the non-planar corrections to this model, which corresponds to the quantum gravity effects. The motivation of this paper is to investigate these non-planar corrections to the PSSY model.

One of the main results of this paper is to show that there is a curious correspondence between the PSSY model and the IOP matrix model \cite{Iizuka:2008eb}, another toy model investigated before as a toy model of proving a black hole. The IOP matrix model is a cousin of the IP model \cite{Iizuka:2008hg} and represents the decay of the correlation function of the probe fundamental field interacting with a matrix degree of freedom describing a black hole. The pros and cons of the IOP matrix model are that it is simpler than the IP model and thus one can solve it in various ways, but the correlator decays only by the power law, not by the exponential. 
As we will show explicitly, the correspondence is seen through the Feynman diagrams of both models. In the IOP matrix model, not only the planar contributions but also the leading non-planar contributions are explicitly calculated in \cite{Iizuka:2008eb}. We show that using the correspondence between the PSSY model and the IOP matrix model, one can evaluate the specific non-planar corrections exactly in the PSSY model. This is the main point of this paper.

However, this is not the end of the story. In the PSSY model, in fact, there are two types of non-planar corrections. The correspondence between the PSSY model and the IOP matrix model enables us to evaluate only one class of non-planar corrections, which involves the diagrams of ``crossing''. On the other hand, the other non-planar corrections are associated with the extra-handle-in-bulk diagrams. For the extra-handle-in-bulk diagrams, there is no associated diagram in the IOP matrix model.  Thus, the correspondence is not completely one-to-one and it does not directly answer all non-planar corrections. Therefore, one needs to do the direct bulk calculation of resummation of such extra-handle-in-bulk diagrams. We leave these extra-handle-in-bulk calculations for future work.

The organization of this short note is as follows. In section \ref{sec:2}, we review both the PSSY model and the IOP matrix model. Section \ref{sec:3} is our main result, where we show there is a correspondence between the PSSY model and the IOP matrix mode, and through that, we evaluate the non-planar corrections involving the diagram of crossing in the PSSY model. In section \ref{sec:conclusion}, we conclude and discuss open issues as well as possible generalizations of our works.

\section{The PSSY model and the IOP matrix model in the planar limit}\label{sec:2}

In this section, the PSSY model (or the West Coast model) and the IOP matrix model are reviewed. We focus on the spectral density of a reduced density matrix in a microcanonical ensemble of the PSSY model and the spectral density of a two-point function of fundamental fields in the IOP matrix model. After reviewing the two models, in Section \ref{sec:3}, we will point out that both spectral densities in the planar limit are represented by the Marchenko-Pastur distribution in random matrix theory and explain how both models correspond to each other.

\subsection{Review of the PSSY model}\label{sec:2.1}
The PSSY model \cite{Penington:2019kki} consists of a black hole in JT gravity with an end-of-the-world (EOW) brane behind the horizon with tension $\mu\ge0$. Its Euclidean action is 
\begin{align}
S=&\;S_{\text{JT}}+\mu\int_{\text{Brane}} ds,\\
S_{\text{JT}}=&\;-\frac{S_0}{4\pi}\left(\int_{\mathcal{M}}\sqrt{g}R+2\int_{\partial M}\sqrt{h}K \right)-\left(\frac{1}{2}\int_\mathcal{M} \sqrt{g}\phi(R+2)+\int_{\partial \mathcal{M}} \sqrt{h}\phi K\right).
\end{align}
We impose the standard asymptotic boundary condition 
\begin{align}
ds^2|_{\partial \mathcal{M}}=\frac{d\tau^2}{z_\epsilon^2}, \;\;\; \phi|_{\partial \mathcal{M}}=\frac{1}{z_\epsilon},
\end{align}
where $\tau$ is the boundary Euclidean time, and $z_\epsilon$ is the near-boundary cutoff. 

Suppose that there are $k$ orthogonal states $|i\rangle_\bold{R}$ of the ``radiation''  system $\bold{R}$, which are entangled with $k$ interior of the EOW brane microstates $|\psi_i\rangle_{\bold{B}}$ of the black hole $\bold{B}$. 
A pure state $|\Psi\rangle$ representing this entanglement is given by
\begin{align}
|\Psi\rangle=\frac{1}{\sqrt{k}}\sum\limits_{i=1}^k\,\ket{\psi_i}_\bold{B}\ket{i}_{\bold{R}},
\end{align}
where the radiation system $\bold{R}$ can be interpreted as the early radiation of an evaporating black hole. The reduced density matrix $\rho_\bold{R}$ and its resolvent $R(\lambda)$ are defined by
\begin{align}
\rho_{\bold{R}} &:=\Tr_\bold{B} \ket{\Psi}\bra{\Psi}=\frac{1}{k}\,\sum\limits_{i,j=1}^k\,\ket{j}\bra{i}_{\bold{R}}\,\braket{\psi_i|\psi_j}_\bold{B},\\
R(\lambda) &:= \sum\limits_{i=1}^k R_{ii}(\lambda)\,,\;\;\;
R_{ij}(\lambda) :=\left( \frac{1}{\lambda \mathds{1} -\rho_\bold{R}}\right)_{ij} \,=\,\frac{1}{\lambda}\,\delta_{ij}+\sum\limits_{n=1}^{\infty}\,\frac{1}{\lambda^{n+1}}\,(\rho_\bold{R}^n)_{ij} \,.
\label{defofRij}
\end{align}

\begin{figure}[t]
\center \includegraphics[width=12cm]{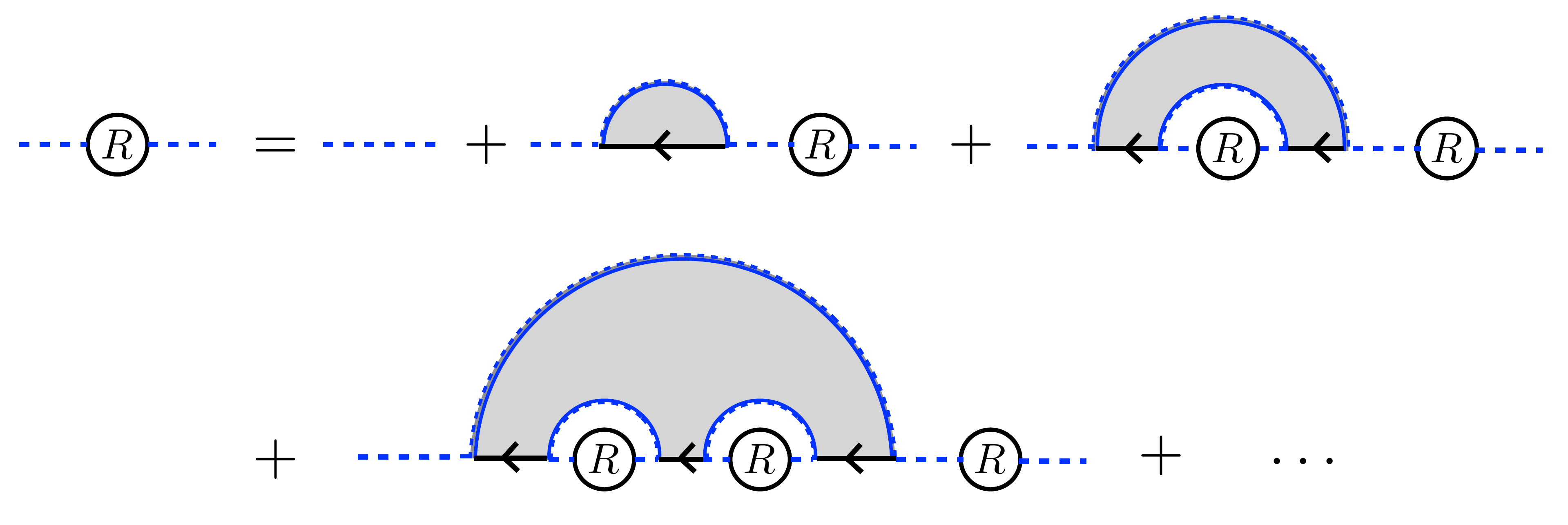}
\caption[]{Schwinger-Dyson equation for the PSSY model in the planar limit.} 
\label{fig:SDPSSY}
\end{figure}

When $e^{\bold{S}}$, which is the dimensions of $\bf{B}$, and $k$, which is the dimension of $\bf{R}$, are large, only planar diagrams are dominant in the Schwinger-Dyson  equation of $R_{ij}(\lambda)$, as Fig.~\ref{fig:SDPSSY}, thus we obtain
\begin{align}
    R_{ij}(\lambda)=\,\frac{1}{\lambda}\,\delta_{ij}+\frac{1}{\lambda}\sum\limits_{n=1}^{\infty}\,\frac{Z_n^{\text{Disk}}}{(k Z_1^{\text{Disk}})^n}\,R(\lambda)^{n-1}R_{ij}(\lambda),\label{SDeq1}
\end{align}
where $\delta_{ij}/\lambda$ is like ``bare propagator'' and $Z_n^{\text{Disk}}$ is the bulk partition function on a disk topology with $n$ asymptotic boundaries represented by the black solid arrows and $n$ blue curved lines for the EOW branes. In a microcanonical ensemble with fixed energy $E$, the ratio of the bulk partition functions is simplified as
\begin{align}
\label{S02E}
\frac{Z_n^{\text{Disk}}}{(Z_1^{\text{Disk}})^n}=e^{-(n-1)\bold{S}}, \;\;\; e^{\bold{S}}:=e^{S_0}\rho_{\text{Disk}}(E)\Delta E, \;\;\; \rho_{\text{Disk}}(E):=\frac{\sinh (2\pi\sqrt{2E})}{2\pi^2},
\end{align}
where $E$ dependence appears through $\bold{S}$, and $\Delta E$ is the width of the microcanonical energy window. Performing the infinite sum in eq.~(\ref{SDeq1}), we obtain
\begin{align}
R(\lambda)^2+\left(\,\frac{e^\textbf{S}-k}{\lambda}-ke^\textbf{S}\,\right)\,R(\lambda)+\dfrac{k^2e^\textbf{S}}{\lambda}\,=0,
   \end{align}
    and a solution of $R(\lambda)$ with the asymptotic behavior $R(\lambda)\to k/\lambda$ at $\lambda \to + \infty$ is
    \begin{align}\label{Rplanar}
    R(\lambda) 
        &= \frac{ k e^\textbf{S}}{2 \lambda } \left(  \left(   e^{-\textbf{S}}- k^{-1}\right)  + \lambda  -   \sqrt{ (\lambda - \lambda_+) (\lambda - \lambda_-)}   \right)   {  \qquad  \left( \mbox{for} \,\,\, \lambda > \lambda_+ \right)},\\
\label{SpectralDensityWest1}
& \qquad \qquad \mbox{where} \quad \lambda_\pm :=\left(k^{-\frac{1}{2}}\pm e^{-\textbf{S}/2} \right)^2.
   \end{align}
$R(\lambda)$ for $\lambda < \lambda_+$ can be obtained by the analytic continuation. 
From the definition of $R(\lambda)$ in \eqref{defofRij}, using 
\begin{align}
\label{SokhotskiPlemelj}
\frac{1}{\lambda + i \epsilon} = \mbox{P}\left( \frac{1}{\lambda} \right) - i \pi \delta(\lambda),
\end{align}
the spectral density $D(\lambda)$ of $\rho_{\bold{R}}$ is given by
\begin{align}
D(\lambda) &=  - \frac{1}{\pi} \Im R(\lambda + i \epsilon)  \notag\\
& =\frac{ke^\textbf{S}}{2\pi\lambda}\sqrt{(\lambda-\lambda_-)(\lambda_+-\lambda)}\theta(\lambda-\lambda_-)\theta(\lambda_+-\lambda)+\left(k-e^{\mathbf{S}}\right)\delta(\lambda)\theta(k-e^{\mathbf{S}})\,,\label{SpectralDensityWest}
\end{align}
where $\theta(\lambda)$ is the Heaviside step function\footnote{From \eqref{SpectralDensityWest1}, we have 
\begin{align}
            R(\lambda)    = \frac{ k e^\textbf{S}}{2 \lambda } \left(  \left(   e^{-\textbf{S}}- k^{-1}\right)  + \lambda + \sqrt{ (\lambda_+ - \lambda) (\lambda_- - \lambda)}   \right)   {  \qquad  \mbox{for} \quad \left(\lambda_+ \ge \lambda_- \ge \lambda \ge 0 \right)}.
\end{align} 
The relative sign in front of square root changes between $\lambda > \lambda_+$ and $0 \le \lambda < \lambda_-$ because we change both of the argument $\theta_+$ and $\theta_-$ by $\pi$ in $\lambda - \lambda_+ := r_+ e^{i \theta_+}$ and $\lambda - \lambda_- := r_- e^{i \theta_-}$. See, for instance, \cite{Lu:2014jua}. Thus, $\lambda = 0$ pole in $R(\lambda)$ gives a Dirac delta function proportional to 
\begin{align}
\frac{{ke^\textbf{S}} }{2} \left(   e^{-\textbf{S}} - k^{-1}  + \sqrt{\lambda_+ \lambda_-}\right) 
=\left(  k - e^\textbf{S} \right) \theta(  k  - e^\textbf{S} ).
\end{align}
}.

One can check that the normalization of $D(\lambda)$ is 
\begin{align}\label{normalizationWest}
\int D(\lambda) d \lambda=k, \quad \int D(\lambda) \lambda \,d \lambda=1.
\end{align}
The first normalization means that the size of $\rho_\bold{R}$ is $k$, and the second normalization means that $\Tr_{\bold{R}}\rho_\bold{R}=1$. $D(\lambda)$ is simplified as
\begin{align}\label{SpectralDensityWestN}
D(\lambda)=\frac{k^2}{2\pi\lambda}\sqrt{\lambda\left(\frac{4}{k}-\lambda\right)}\theta(\lambda)\theta\left(\frac{4}{k}-\lambda\right) \,, \quad \mbox{when} \quad k=e^\textbf{S} \,.
\end{align}

Using \eqref{SpectralDensityWest1} and \eqref{SpectralDensityWest}, the entanglement entropy $S_\bold{R}$ of the auxiliary system $\bold{R}$ can be calculated as
\begin{align}\label{SE}
S_{\bold{R}}=&\;-\int d\lambda D(\lambda) \lambda\log \lambda\notag\\
=&\;-\frac{ke^\textbf{S}}{2\pi}\int _{\lambda_-}^{\lambda_+}d\lambda \sqrt{(\lambda-\lambda_-)(\lambda_+-\lambda)}\log \lambda.
\end{align}
This integral can be computed exactly, and the result is
\begin{align}
    S_{\bold{R}}=\log m -\frac{m}{2n}, \;\;\; m:=\min \{k, e^\textbf{S}\}, \;\;\; n:=\max \{k, e^\textbf{S}\},
\end{align}
which perfectly matches the Page's result for $n\ge m\gg1$ \cite{Page:1993df}. If $k=e^\textbf{S}$, the entanglement entropy is
\begin{align}\label{SE2}
    S_{\bold{R}}=\log k -\frac{1}{2}  \,, \quad \mbox{if} \quad k=e^\textbf{S}.
\end{align}

\subsection{Review of the IOP matrix model}

The IOP matrix model \cite{Iizuka:2008eb} is a matrix model given by the following Hamiltonian 
\begin{align}
H_{IOP}=m A_{ij}^\dagger A_{ji}+M a_i^\dagger a_i+H_{int}, \;\;\; H_{int}=h a_{i}^\dagger a_lA_{ij}^\dagger A_{jl}, 
\end{align}
where the sum of subscripts is taken from $1$ to $N$. Here, $a_i$ is the annihilation operator for a harmonic oscillator in the fundamental of $U(N)$, and $A_{ij}$ is the annihilation operator for an oscillator in the adjoint. This matrix model was introduced as a toy model of the gauge dual of an AdS black hole, where the adjoint fields can be interpreted as background $N$ D0-branes for the black hole, and the fundamental fields can be interpreted as strings stretched from a probe D0-brane.

To solve the spectrum density analytically in the large $N$ limit with fixed 't Hooft coupling $\lambda_{\text{'t Hooft}}:= hN$, we also take the large $M$ limit $M\gg m$ and $M\gg T$ so that $a_i^\dagger a_i\sim0$ in the thermal ensemble at finite temperature $T$.
We consider the following time-ordered Green’s functions at finite temperature
\begin{align}
e^{i M t} \Big{\langle} \mbox{T} \, a_i(t)\, a_j^\dagger(0) \Big{\rangle}_T &=:  G(t)\delta_{ij},\\
\Big\langle \mbox{T}  A_{i j}(t) A_{k l}^{\dagger}(0)\Big\rangle_T&=:L(t) \delta_{i l} \delta_{j k}.
\end{align}
With the Fourier transformation $\tilde{f}(\omega)=\int dt \, e^{i\omega t} f(t)$, free thermal propagators in frequency space are given by
\begin{align}
\tilde{G}_0(\omega) & = \frac{i}{\omega + i \epsilon},\label{freeG0}\\
\tilde{L}_0 (\omega) &= \frac{i}{1-y}\left(\frac{1}{\omega-m + i \epsilon} - \frac{y}{\omega-m - i \epsilon}\right), \;\;\; y:=e^{-m/T},\label{L0}
\end{align}
where $\tilde{G}_0(\omega)$ does not depend on $T$ in the large $M$ limit, and $\tilde{L}(\omega)$ in the large $N$ limit becomes the free propagator $\tilde{L}_0(\omega)$ since the backreaction from the fundamental is suppressed by $1/N$.

\begin{figure}[t]
\center \includegraphics[width=12cm]{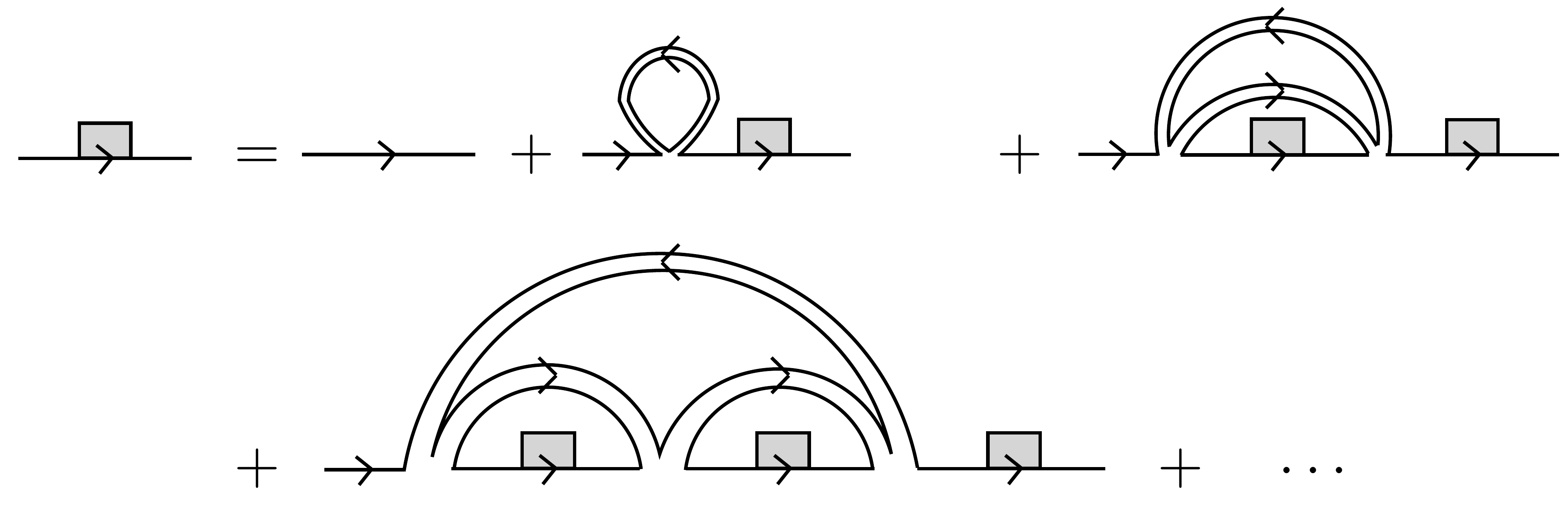}
\caption[]{Schwinger-Dyson equation for the IOP matrix model in the planar limit.} 
\label{fig:SDIOP}
\end{figure}
In the limit where $N$ and $M$ are large, the Schwinger-Dyson equation of $\tilde{G}(\omega)$ is shown in Fig.~\ref{fig:SDIOP}, which has the same graphical structure as the Schwinger-Dyson equation of $R(\lambda)$ in the PSSY model. See Figure 2 of \cite{Iizuka:2008eb} and Figure (2.25) of \cite{Penington:2019kki} as well. 
The Schwinger-Dyson equation of $\tilde{G}(\omega)$ is given by
\begin{align}
\tilde G(\omega) = \tilde G_0(\omega) +  y\tilde{G}_0(\omega) \tilde G(\omega)
\sum_{n=0}^\infty \left(\frac{-i \lambda_{\text{'t Hooft}}}{1-y} \right)^{n+1} \tilde G(\omega)^n.
\end{align}
Performing the infinite sum, we obtain
\begin{align}
\tilde G(\omega)^2-i(1-y)\left(\frac{1}{\omega}+\frac{1}{\lambda_{\text{'t Hooft}}}\right)\tilde G(\omega)-\frac{1-y}{\omega\lambda_{\text{'t Hooft}}}=0,
\end{align}
and its solution is
\begin{align}
\tilde G(\omega) 
 &= \frac{i}{2 \omega} \frac{1-y}{\lambda_{\text{'t Hooft}} }\left( \frac{\lambda_{\text{'t Hooft}}}{1-y} (1-y) + \omega  - \sqrt{\left(\omega - \omega_+ \right)
\left( \omega - \omega_- \right)} \right)\label{dressedGX}  \,\,\, (\mbox{for} \,\,\, \omega > \omega_+) \,,   \\
& \qquad \qquad \mbox{where} \quad \omega_{\pm}  := \frac{\lambda_{\text{'t Hooft}}}{1-y}  \left( 1 \pm  \sqrt{y} \right)^2  \ge 0 \,,
\end{align}
where $0 \le y \le1$ and we take the branch such that $\tilde G(\omega)$ at $\omega \to +\infty$ becomes the free propagator given by \eqref{freeG0}. $\tilde G(\omega)$ for $\omega < \omega_+$ can be obtained by the analytic continuation. 
Again using \eqref{SokhotskiPlemelj} and \eqref{dressedGX},  the spectral density $F(\omega)$ of the two-point function of fundamental fields is obtained as\footnote{From \eqref{dressedGX}, we have 
\begin{align}
\tilde G(\omega) 
 &= \frac{i}{2 \omega} \frac{1-y}{\lambda_{\text{'t Hooft}} }\left( \frac{\lambda_{\text{'t Hooft}}}{1-y} (1-y) + \omega  + \sqrt{\left(\omega_+ - \omega \right)
\left( \omega_- - \omega \right)} \right),\label{dressedGX2}  \,\,\, (\mbox{for} \,\,\, \omega_+ \ge \omega_- \ge \omega > 0) \,.  
\end{align}
Again, the relative sign in front of the square root changes between $\omega > \omega_+$ and $0 \le \omega < \omega_-$. 
Thus, $\omega \to 0$ pole gives a Dirac delta function proportional to 
\begin{align}
&\frac{1}{2} \frac{1-y}{\lambda_{\text{'t Hooft}} }\left( \frac{\lambda_{\text{'t Hooft}}}{1-y} (1-y)  + \sqrt{  \omega_+ 
 \omega_- } \right) 
 = (1-y)\theta( 1-y ). 
\end{align}} 
\begin{align}
F(\omega) &=  \frac{1}{\pi} \Re \tilde G(\omega + i \epsilon)\\
&= \frac{1}{2 \pi \omega} \frac{1-y}{\lambda_{\text{'t Hooft}}}\sqrt{(\omega-\omega_-)(\omega_+-\omega)}\theta(\omega-\omega_-)\theta(\omega_+-\omega) \nonumber \\
& \qquad +(1-y)  \theta(1-y)  \delta(\omega)  \,. \label{SpectralDensityIOP}
\end{align}
Note that our convention of the propagators includes a factor $i$ in the numerator as seen in eq.~(\ref{freeG0}). 
$F(\omega)$ is normalized as 
\begin{align}
\label{IOPnorm}
\int  F(\omega) d\omega= 1, \;\;\; \int F(\omega)\omega d\omega=\frac{y\lambda_{\text{'t Hooft}}}{1-y}.
\end{align}

\section{The PSSY model and the IOP matrix model correspondence}\label{sec:3}

\subsection{Feynman diagram correspondence between the PSSY model and the IOP matrix model}
As seen in Fig.~\ref{fig:SDPSSY} and \ref{fig:SDIOP}, in the planar limit, the Schwinger-Dyson equations in the PSSY model and the IOP matrix model have the same graphical structure. From now on, we elaborate on the correspondence of each diagram.

\begin{figure}[t]
\center \includegraphics[width=12cm]{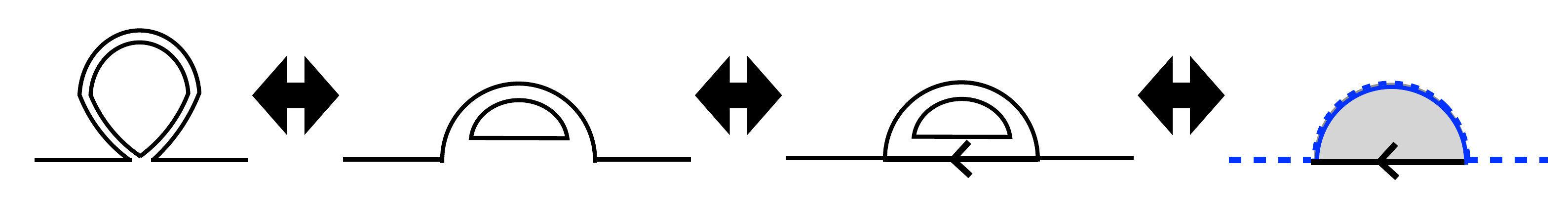}
\caption[]{The prescription to change the IOP vertex (left) to the PSSY vertex (right).} 
\label{fig:prescription}
\end{figure}

From diagrams in the IOP matrix model, one can uniquely construct the corresponding diagrams in the PSSY model and vice versa. The Feynman diagram correspondence can be obtained by the following prescription. See Fig.~\ref{fig:prescription}.

\begin{enumerate}
    \item Extend vertices in the IOP matrix model horizontally and draw straight lines with horizontal arrows from right to left. These arrows represent the asymptotic boundaries with the time direction from the ket to the bra in the PSSY model.  
    \item Rewrite the adjoint correlators in the IOP matrix model as blue solid curves in the PSSY model. These blue solid curves correspond to EOW branes in the PSSY model.
    \item Fill in regions above the right-to-left horizontal arrows corresponding to asymptotic boundaries with a gray shadow. These shaded regions correspond to bulk geometries in the PSSY model. 
\end{enumerate}

\begin{figure}
\center \includegraphics[width=12cm]{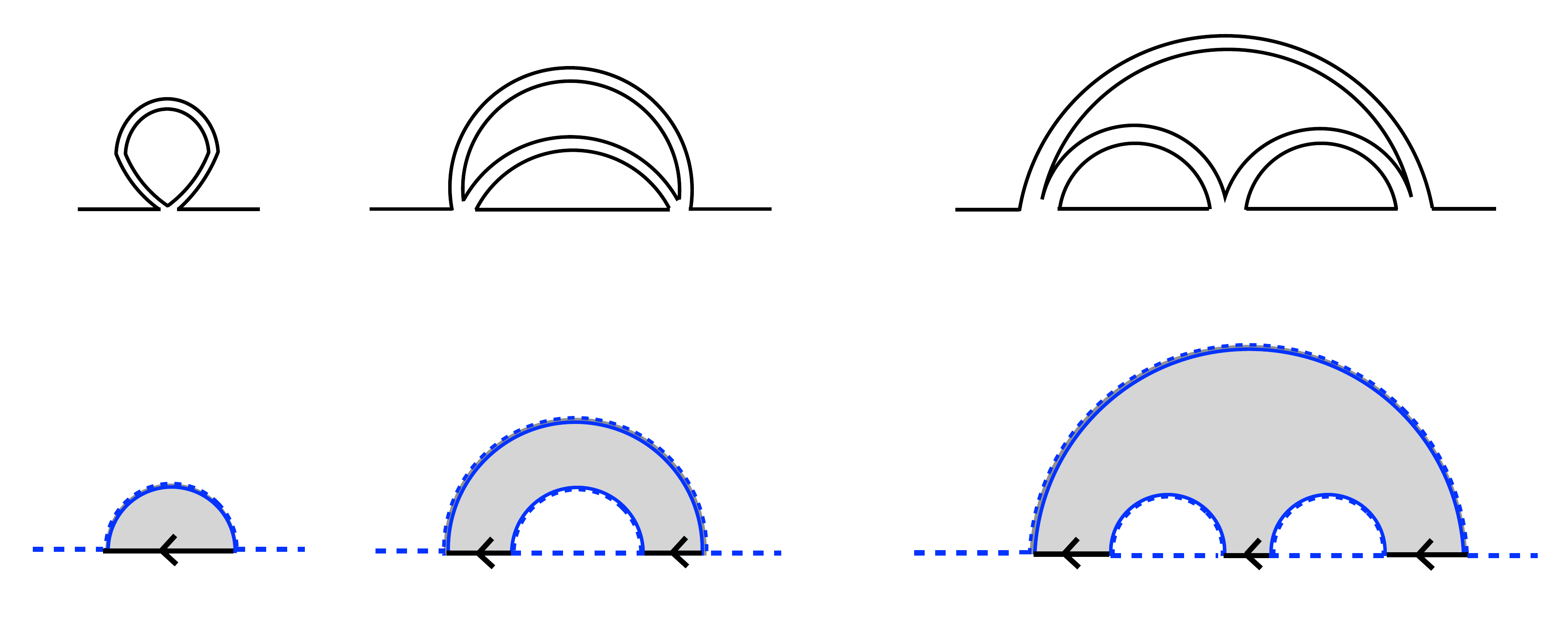}
\caption[]{Corresponding planar diagrams in the IOP matrix model (upper diagrams) and the PSSY model (lower diagrams).} 
\label{fig:planardiagrams}
\end{figure}

Figure \ref{fig:planardiagrams} shows examples of corresponding planar diagrams, where we omit arrows in the IOP matrix model for easier comparison. Due to the correspondence between these diagrams, there is also the correspondence between the solutions of the Schwinger-Dyson equations in the planar limit. The correspondence between parameters in both models is examined in the next subsection. 

\begin{figure}
\center \includegraphics[width=12cm]{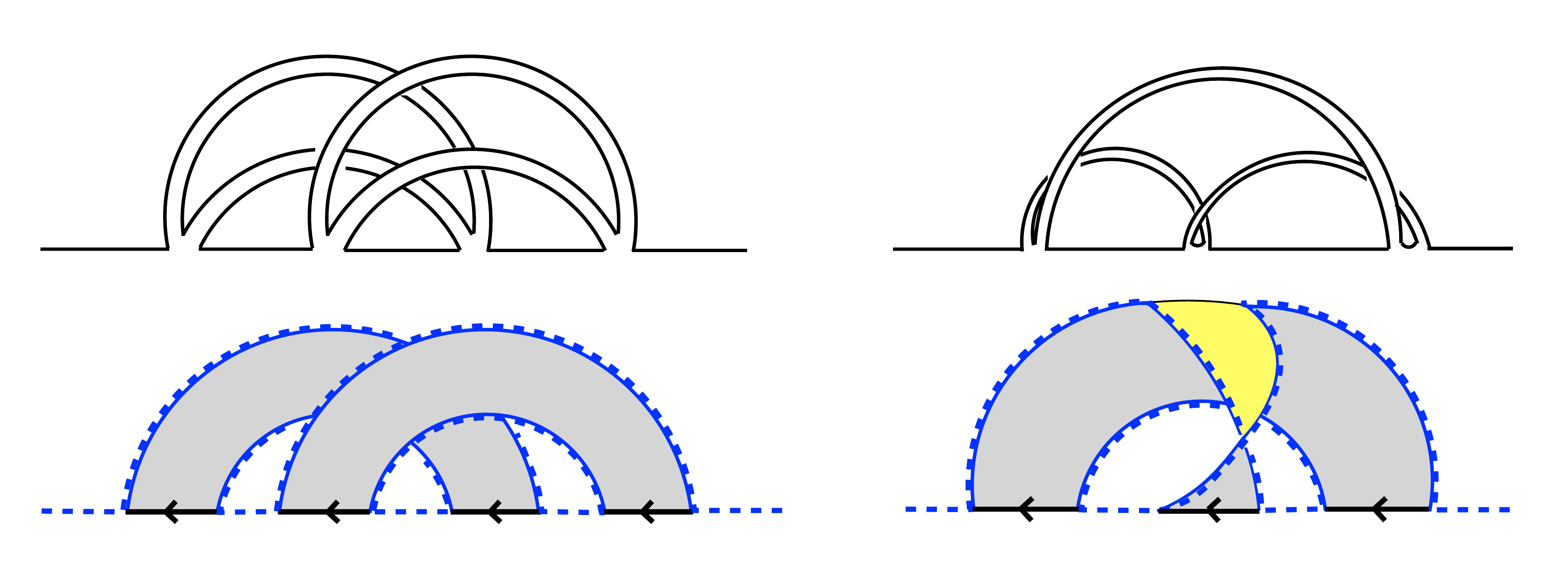}
\caption[]{Corresponding non-planar diagrams in the IOP matrix model (upper diagrams) and the PSSY model (lower diagrams).} 
\label{fig:nonplanardiagrams}
\end{figure}

Due to the correspondence, there is one-to-one Feynman diagram correspondence between the IOP matrix model Feynman diagrams and the PSSY model Feynman diagrams. Thus, the correspondence goes beyond the planar limit. For example, Figure \ref{fig:nonplanardiagrams} shows examples of corresponding non-planar diagrams. From the perspective of the PSSY model, the left figure includes two bulk geometries with a crossing, and the right figure includes a twisted bulk geometry that is anchored to the asymptotic boundaries. 

\begin{figure}
\center \includegraphics[width=12cm]{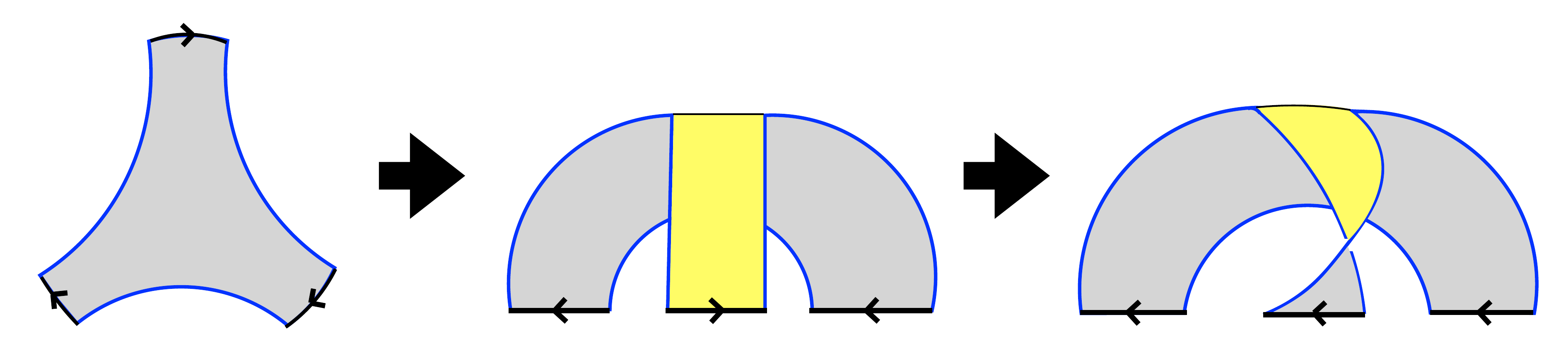}
\caption[]{How to construct a twisted bulk geometry (right diagram) by twisting a bulk geometry for $Z_3^{\text{Disk}}$ (left diagram). The yellow-shaded surface represents the reverse side of the gray surface.} 
\label{fig:twistedgeometry}
\end{figure}

Let us look into a little more on the twisted bulk geometry in Figure \ref{fig:nonplanardiagrams}. We can construct this twisted bulk geometry from a bulk geometry for $Z_3^{\text{Disk}}$ as follows. First, prepare the bulk geometry for $Z_3^{\text{Disk}}$ shown at left in Figure \ref{fig:twistedgeometry}. Next, fold the top part downward so that the yellow reverse side is visible as shown in the middle figure. Finally, twist the folded part so that the middle arrow is facing left as shown in the right figure.

\begin{figure}
\center \includegraphics[width=12cm]{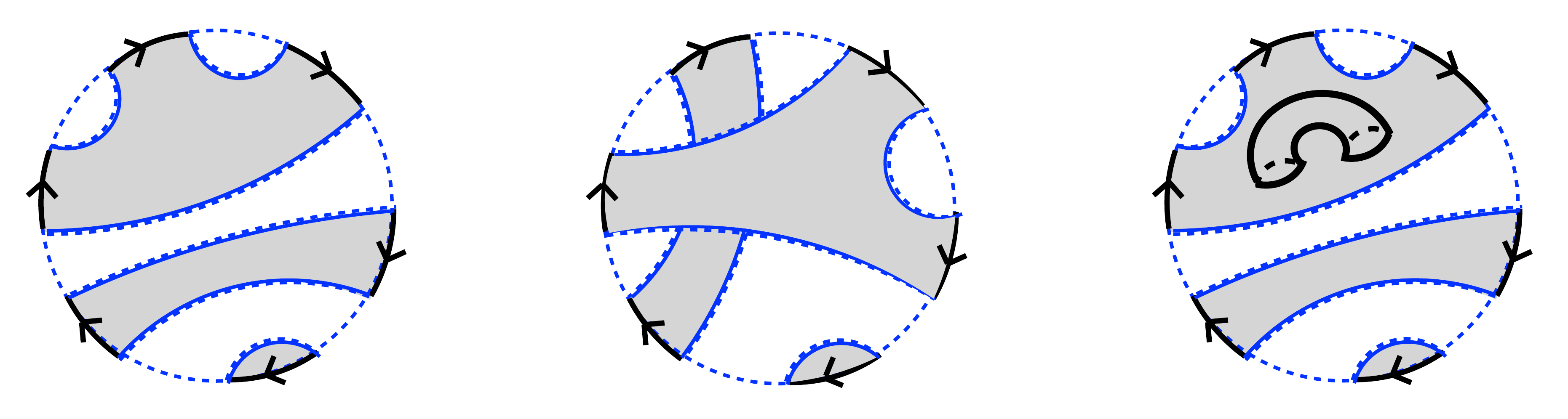}
\caption[]{Three examples of bulk geometries that contribute to $\Tr (\rho^6_{\bold{R}})$. The left figure is a planar geometry, the middle figure is a non-planar geometry with a crossing, and the right figure is a non-planar geometry with an extra handle in bulk.} 
\label{fig:IOPWormhole}
\end{figure}

Following our prescription in reverse, we can also construct the corresponding diagrams in the IOP matrix model from the ones in the PSSY model. However, note that not all bulk geometries in the PSSY model correspond to diagrams in the IOP matrix model. To see this point, for instance, let us consider three examples of bulk geometries that contribute to $\Tr (\rho^6_{\bold{R}})$ as shown in Figure \ref{fig:IOPWormhole}. In the IOP matrix model, there are diagrams corresponding to the planar one and the non-planar one with a crossing such as the left and middle geometries in Figure \ref{fig:IOPWormhole}, respectively. However, there is no diagram in the IOP matrix model for the non-planar geometry with an extra handle such as the right geometry where non-planar effects are due to the extra handle in bulk, not crossings.

These clearly show that the correspondence works as long as we neglect the extra-handle-in-bulk diagrams.
Thus in this paper, using the PSSY and the IOP model correspondence, in Subsection \ref{subsec:3.3}, we calculate the exact non-planar effects associated with a crossing as the middle figure in Fig.~\ref{fig:IOPWormhole}. 

\subsection{Parameter correspondence between the PSSY model and the IOP matrix model}
Given the one-to-one Feynman diagram correspondence, it is straightforward to read off the parameter correspondence between them. One can observe that the spectral densities $D(\lambda)$ in \eqref{SpectralDensityWest} and $F(\omega)$ \eqref{SpectralDensityIOP} have the same structures. In fact, after rescaling, the spectral density of the IOP matrix model at infinite temperature limit agrees with the spectral density $D(\lambda)$ (\ref{SpectralDensityWestN}) of the PSSY model when $k=e^{\bold{S}}$.

The reason why we need to take an infinite temperature in the IOP matrix model is as follows. 
In the propagator $\tilde{L}_0(\omega)$ (\ref{L0}) of the IOP matrix model, there is a difference of a factor $y$ between the two terms. See eq.~(3.1) of \cite{Iizuka:2008eb} as well. However, there is no such difference in the PSSY model side. To eliminate this difference, we need to take the following infinite temperature limit
\begin{align}\label{infiniteT}
y=e^{-m/T}\to1  \quad \text{and}  \quad \lambda_{\text{'t Hooft}}\to0 \, \quad \text{with} \;\,\lambda_y:=\frac{\lambda_{\text{'t Hooft}}}{1-y} = \mbox{fixed}.
\end{align}
In this infinite temperature limit, the spectral density $F(\omega)$ (\ref{SpectralDensityIOP}) becomes
\begin{align}
\label{limitIOPF}
F(\omega)
=&\; \frac{1}{2 \pi \omega\lambda_{y}} \sqrt{\omega(4\lambda_y-\omega)}\theta(\omega)\theta(4\lambda_y-\omega).
\end{align}
Since there is a correspondence between the Feynman diagrams in the IOP matrix model and the PSSY model, this $F(\omega)$ should correspond to $D(\lambda)$ up to some normalization. 

Let us compare $D(\lambda)$ in the PSSY model given by \eqref{SpectralDensityWest} and $F(\omega)$ in the IOP matrix model given by \eqref{SpectralDensityIOP}. Then it is clear that under the $y \to 1$ limit, one needs $k=e^{\bold{S}}$ limit for the correspondence to work\footnote{We will later see in the discussion section that in order to go beyond $k=e^{\bold{S}}$ limit, one needs to consider a rectangular model. As long as we are considering a square matrix in the IOP model, one has to take $k=e^{\bold{S}}$ limit for the correspondence to the PSSY model to work. Note that even in the rectangular model, one always needs $y\to 1$ limit in the IOP model for the correspondence to the PSSY model for $k \neq e^{\bold{S}}$.}. Thus we focus on this limit.   
Furthermore, in order to take into account the normalization difference between $F(\omega)$ in the IOP matrix model \eqref{IOPnorm} and $D(\lambda)$ in the PSSY model \eqref{normalizationWest}, we divide $D(\lambda)$ in \eqref{SpectralDensityWestN} by $k$ as 
\begin{align}
\label{limitPSSYD}
\frac{1}{k} D(\lambda)=\frac{k}{2\pi \lambda}\sqrt{\lambda \left(\frac{4}{k}-\lambda \right)}\theta(\lambda)\theta\left(\frac{4}{k}-\lambda \right)  \,, \quad \mbox{when} \quad k=e^\textbf{S} \,.
\end{align}
Let's compare \eqref{limitIOPF} and \eqref{limitPSSYD}. One might naively think that $\omega$ in the IOP matrix model corresponds to $\lambda$ in the PSSY model. However, this cannot be true since their dimensions do not match. To make $\omega$ dimensionless and also to match the parameter range in $\theta$-functions, we define 
\begin{align}
\tilde{\omega}:=\frac{\omega}{\lambda_y k} \quad \mbox{such that} \quad
0 \le \omega \le 4 \lambda_y  \, \Leftrightarrow \, 0 \le \tilde{\omega} \le \frac{4}{k}.
\label{deftildeomega}
\end{align}
Then, we can define 
\begin{align}\label{rescaling}
\tilde{F}(\tilde{\omega}):=\lambda_y k F(\omega)   \quad \mbox{such that} \quad
\int d \tilde{\omega} \tilde{F}(\tilde{\omega}) = \int d \omega F(\omega) = 1.
\end{align}
Thus, we obtain 
\begin{align}
\label{rescaledtildeF}
\tilde{F}(\tilde{\omega}) = \frac{k}{2 \pi \tilde{\omega} }\sqrt{\tilde{\omega} \left( \frac{4}{k} - \tilde{\omega}\right)  }\theta \left(\tilde{\omega} \right) \theta \left(\frac{4}{k} - \tilde{\omega} \right).
\end{align}
It is then clear that there is a parameter correspondence between the two models as follows
\begin{align}
\label{id1}
&\tilde{\omega} \, (\mbox{IOP}) \, \leftrightarrow \, \lambda \,(\mbox{PSSY})\,, \quad \tilde{F}(\tilde{\omega}) \, (\mbox{IOP}) \, \leftrightarrow \, \frac{1}{k} D(\lambda) \,(\mbox{PSSY})  \,, \\
& \qquad \qquad \qquad N  \, (\mbox{IOP}) \, \leftrightarrow \,  k=e^{\bold{S}}  \,(\mbox{PSSY})\,
\label{id2}
\end{align}
at $y\to 1$ limit. Note that, for the planar limit,  we consider the large $N$ limit in the IOP matrix model and the large $k$, $e^{\bold{S}}$ limit in the PSSY model, and they correspond as \eqref{id2}.

Let us investigate the correspondence in more detail. In the PSSY model, the spectral density $D(\lambda)$ is computed from the resolvent $R(\lambda)$
\begin{align}
R(\lambda)=\Tr \frac{1}{\lambda \mathds{1} -\rho_\bold{R}}=\sum_{l=1}^k\bra{l}_{\bold{R}}\frac{1}{\lambda \mathds{1} -\rho_\bold{R}}\ket{l}_{\bold{R}}, \\ 
\mbox{where} \quad \rho_\bold{R}=\frac{1}{k}\,\sum\limits_{i,j=1}^k\,\ket{j}\bra{i}_{\bold{R}}\,\braket{\psi_i|\psi_j}_\bold{B}.\label{Rl}
\end{align}
In the IOP matrix model, the two-point function $\tilde G (\omega)$ in the large $M$ limit can be expressed as
\begin{align}
& N\tilde G (\omega)=\sum_{l=1}^N\Big{\langle}a_l \frac{i}{\omega\mathds{1}-H_{int}}a^\dagger_l \Big{\rangle}_T \nonumber \\ 
& \implies \quad - i \lambda_yN^2\tilde G (\omega)=\sum_{l=1}^N\Big{\langle}a_l \frac{1}{\frac{\omega}{\lambda_yN}\mathds{1}-\frac{1-y}{N^2} a_j^\dagger a_iA^\dagger_{jk} A_{ki}} a^\dagger_l\Big{\rangle}_T.\label{Go}
\end{align}
Since we take $M$ to be large so that the number of fundamental fields is always one in the evaluation of $\tilde G (\omega)$, we can treat $a_l^\dagger \ket{v}$ as an $N$-dimensional one-particle excited state basis, where $\ket{v}$ is the ground state for the fundamental field. 
Comparing eqs.~ (\ref{Rl}) and (\ref{Go}), we obtain the following additional relationships\footnote{To be precise, since the trace of a matrix is invariant under the transformation of a basis, there is the ambiguity of a unitary matrix $U$ in the correspondence \eqref{id3} as
\begin{align}\label{ambiguity1}
 \qquad \frac{H_{int}}{\lambda_yN}  \,\,\,\,(\mbox{IOP})   \, & \leftrightarrow \,\,\,\,  U\rho_\bold{R}U^\dagger  \,\,\,\, (\mbox{PSSY}) \,.
\end{align}}
\begin{align}
\label{id3}
 \qquad \frac{H_{int}}{\lambda_yN} = \frac{1-y}{N^2} a_j^\dagger a_iA^\dagger_{jk} A_{ki}  \,(\mbox{IOP})   \, & \leftrightarrow \,\,\,\,  \rho_\bold{R}  \,\,\,\, (\mbox{PSSY}) \,, \\
 -i\lambda_yN^2\tilde G (\omega)  \, \,\, (\mbox{IOP})    \,  &\leftrightarrow \, \,R(\lambda)  \, (\mbox{PSSY})  \,. 
\end{align}
in addition to the parameter correspondence given by \eqref{id1} and \eqref{id2}. 
In \eqref{id3}, naively one might wonder if this term vanishes in the $y\to 1$ limit. However, the adjoint propagator is also proportional to $1/(1-y)$ as seen in \eqref{L0}, thus this is a well-defined limit even in $y\to 1$.

Furthermore, $\ket{l}_\bold{R}$, which forms an orthonormal basis for the radiation Hilbert space, corresponds to the one-fundamental excited state $a_l^\dagger \ket{v}$ that is again orthogonal.  
Given these correspondences, one can also see the relationship 
\begin{align}
&\text{Random ensemble average of } \braket{\psi_i|\psi_j}_\bold{B} \, (\mbox{PSSY}) \nonumber \\
 &\leftrightarrow \, \, \text{ Expectation value of } \frac{1-y}{N}A^\dagger_{jl}A_{li} \, \,\, (\mbox{IOP}) .
\end{align}
In the PSSY model, the Gaussian random property of $\braket{\psi_i|\psi_j}_\bold{B}$ is crucial for connected wormhole contributions. From the viewpoint of the IOP matrix model, this Gaussian randomness comes from the fact that the adjoint fields $A^\dagger$ behave like Gaussian free fields. In random matrix theory, the spectral density $D(\lambda)$ (\ref{SpectralDensityWest}) up to the normalization is known as the Marchenko-Pastur distribution \cite{Pastur:1967zca}, see, for instance, \cite{Muck:2024fpb}. The reason why the Marchenko-Pastur distribution appears is that $\braket{\psi_i|\psi_j}_\bold{B}$ in the PSSY model can be interpreted as a Gram matrix \cite{Balasubramanian:2022gmo, Balasubramanian:2022lnw, Climent:2024trz} and $H_{int}$ in the IOP matrix model is proportional to $A^\dagger A$.

\subsection{Non-planar correction of the entanglement entropy in the PSSY model via the IOP matrix model correspondence}\label{subsec:3.3}

Non-planar $1/N^2$ correction of the two-point function $\tilde G(\omega)$ in the IOP matrix model was computed by \cite{Iizuka:2008eb}. By using the PSSY model and the IOP matrix model correspondence, it is straightforward to obtain the non-planar $1/k^2$ correction of the reduced density matrix and its von Neumann entropy in the PSSY model. 
Especially, the spectral density $D(\lambda)$ (\ref{SpectralDensityWestN}) in the PSSY model and the rescaled one $\tilde{F}(\tilde{\omega})$ (\ref{rescaledtildeF}) in the IOP matrix model corresponds in the planar limit. Then the non-planar $1/N^2$ correction of the von Neumann entropy in the IOP matrix model would be a part of the non-planar $1/k^2$ correction of the entanglement entropy in the PSSY model.

The non-planar $1/N^2$ correction of $\tilde G(\omega)$ is calculated in \cite{Iizuka:2008eb}, and it is 
\begin{align}
\tilde{G}(\omega) & = \tilde{G}^{(0)} (\omega) + { 1 \over N^2} \tilde{G}^{(1)} (\omega)+ {\cal{O}}\left({1 \over N^4}\right) \,,\\
\tilde{G}^{(0)} (\omega) & = \frac{i}{2 \lambda_y}\left(1-\sqrt{1-\frac{4 \lambda_y}{\omega}}\right)\,, \\  
x_0  & := -i \lambda_y \tilde{G}^{(0)} (\omega)=\frac{1}{2}\left(1-\sqrt{1-\frac{4 \lambda_y}{\omega}}\right) \,, \\
\tilde{G}^{(1)}(\omega) & =
\frac {i x_0^3  (1 -  x_0)^4  }
 {  (1 - 2 x_0 )^4   (\omega (1-x_0)^2 - \lambda_y)}=\frac{i \lambda_y^3}{(\omega - 4 \lambda_y)^{5/2}\omega^{3/2}}\,. \label{G1}
 \end{align}
By using this result, we obtain the non-planar $1/N^2$ correction of the spectral density 
\begin{align}
F(\omega) & =F^{(0)} (\omega) + { 1 \over N^2} F^{(1)} (\omega)+ {\cal{O}}\left({1 \over N^4}\right),\\
F^{(0)} (\omega) & = \frac{1}{\pi} \Re \tilde G^{(0)}(\omega) 
=\frac{\sqrt{\omega(4\lambda_y-\omega)}}{2\pi \lambda_y \omega}\theta(\omega)\theta\left(4 \lambda_y -\omega\right),\\
F^{(1)} (\omega) & =\frac{1}{\pi} \Re \tilde G^{(1)}(\omega)=\frac{\lambda_y^3}{\pi \omega^{3/2}(4 \lambda_y -\omega)^{5/2} }\theta(\omega)\theta\left(4 \lambda_y -\omega\right).
\label{1/N2inIOP}
\end{align}
Since $\tilde G^{(1)}(\omega)$ (\ref{G1}) is a rational function of $x_0$ and $\omega$, $F^{(1)}(\omega)$ has branch points at $\omega=0,4 \lambda_y$ that are the same branch points of $F^{(0)}(\omega)$. This property comes from the fact that the perturbation equation determining $\tilde{G}^{(1)}(\omega)$ is written in terms of $\tilde{G}^{(0)}(\omega)$. Note that even though the branch points of $F^{(0)}(\omega)$ and $F^{(1)}(\omega)$ are the same, $F^{(1)}(\omega)$ is singular than $F^{(0)}(\omega)$. 

Given the correspondence we discussed in the previous subsection, we can read off the $1/k^2$ corrections in the PSSY model from \eqref{deftildeomega}, \eqref{rescaling}, \eqref{id1}, \eqref{id2}, and \eqref{1/N2inIOP} as 
\begin{align}
\frac{1}{k} D^{(0)}(\lambda)& =\frac{k}{2\pi \lambda}\sqrt{\lambda \left(\frac{4}{k}-\lambda \right)}\theta(\lambda)\theta\left(\frac{4}{k}-\lambda \right)  \,, \\
\frac{1}{k} D^{(1)}(\lambda)&= \frac{1}{\pi k^3} \frac{1}{\lambda^{3/2} (\frac{4}{k} - \lambda)^{5/2}}\theta(\lambda)\theta\left(\frac{4}{k}-\lambda \right)  \,, 
\end{align}
when $k=e^\textbf{S}$. Here $D^{(0)}(\lambda)$ and $D^{(1)}(\lambda)$ are the same order since 
\begin{align}
\frac{1}{k} D^{(0)}(\lambda) = \order{k} \,, \quad 
\frac{1}{k} D^{(1)}(\lambda) = \order{k} \, \quad \mbox{for $\lambda \sim \frac{1}{k}$}.
\end{align}
With this, one can calculate the entanglement entropy for the radiation $S_{\bold{R}}$ as 
\begin{align}\label{vnentropy}
S_{\bold{R}} &:=-\int D(\lambda) \lambda  \log \lambda \notag\\
&=-\int \left( D^{(0)}(\lambda) + \frac{1}{k^2}  D^{(1)}(\lambda)  + \order{1 \over k^4}  \right)  \lambda  \log \lambda. 
\end{align}

The leading term can be evaluated as
\begin{align}\label{leadingterm}
-\int   D^{(0)}(\lambda) \lambda  \log \lambda 
=- \frac{k^2}{2\pi} \int_0^{4/k}\sqrt{\lambda \left({4 \over k}-\lambda \right)}\log \lambda  \, d \lambda 
=\log k -\frac{1}{2},
\end{align}
which agrees with eq.~(\ref{SE2}). 
The subleading term is 
\begin{align}
-\frac{1}{k^2}\int D^{(1)}(\lambda)  \lambda  \log \lambda \, d \lambda  
&= - \frac{1}{\pi k^4} \int_0^{4/k}  \frac{\lambda}{\lambda^{3/2} \left(\frac{4}{k} - \lambda \right)^{5/2}} \log \lambda \, d \lambda 
=\frac{C}{k^2}, \\
\mbox{where} \quad C &:= -\frac{1}{\pi}\int_0^4\frac{x^2}{\left(x(4-x)\right)^{5/2}}\log \left( \frac{x}{k} \right) d x, 
\end{align}
where we change the integration variable $\lambda = x/k$. 

$C$ does not converge due to more singular nature of $D^{(1)}(\lambda)$ than $D^{(0)}(\lambda)$. To regularize this integral, we introduce a small cutoff $\epsilon$ so that $C$ is regularized as 
\begin{align}\label{Cepsilon}
C_\epsilon &:=  -\frac{1}{\pi} \int_{0}^{4-\epsilon} \frac{x^2}{\left(x(4-x)\right)^{5/2}}\log \left( \frac{x}{k} \right) d x \\
& = \frac{\log \frac{k}{4}}{3 \pi  \epsilon^{3/2}}+\frac{\log \frac{k}{4}+2}{8 \pi 
   \sqrt{\epsilon}}+\frac{1}{12}+\order{\epsilon^{1/2}}.
\end{align} 
The first and second terms are divergent but they are regulator dependent.  On the other hand, $1/12$ is a regulator-independent one. Thus, we focus on this $1/12$ by subtracting the UV divergent terms\footnote{One might be able to justify this argument along the line of ``renormalized entanglement entropy'' of \cite{Liu:2012eea}.}.

Therefore, after subtracting the regulator-dependent divergent terms at $\epsilon\to0$, we obtain a finite result
\begin{align}\label{IOPSubleadingVNE}
S_{\bold{R}} =  \log k -\frac{1}{2}+\frac{1}{12 k^2}+{\cal{O}}\left({1 \over k^4}\right) \,, \quad \mbox{when} \quad k = e^{\bold{S}}.
\end{align}

Since the leading term (\ref{leadingterm}) in $S_{\bold{R}}$ agrees with $S_\bold{R}$ (\ref{SE2}) in the planar limit, we expect that the sub-leading term in $S_{\bold{R}}$ (\ref{IOPSubleadingVNE}) corresponds to a part of non-planar $1/k^2$ corrections of $S_\bold{R}$ in the PSSY model. As shown in Fig.~\ref{fig:IOPWormhole}, non-planar corrections of $S_\bold{R}$ in the PSSY model come from non-planar diagrams with extra-handle-in-bulk and crossings. We expect that the sub-leading term in $S_{\bold{R}}$ (\ref{IOPSubleadingVNE}) corresponds to the non-planar correction of $S_\bold{R}$ from crossings, not extra-handle-in-bulk.

\section{Short conclusion and discussions}
\label{sec:conclusion}

As mentioned in Subsection \ref{sec:2.1}, the entanglement entropy in the PSSY model and the one of a random pure state coincide with each other in the planar limit for large Hilbert space dimensions. We expect that non-planar corrections in the PSSY model and the IOP matrix model have some connection to Page's conjecture \cite{Page:1993df, Foong:1994eja, PhysRevE.52.5653, Sen:1996ph} on the entanglement entropy of a random pure state for general Hilbert space dimensions. 

Page's conjecture on the entanglement entropy $S_R$ of a random pure state \cite{Page:1993df}, which was probed by \cite{Foong:1994eja, PhysRevE.52.5653, Sen:1996ph}, is
\begin{align}
    S_R=\sum_{k=n+1}^{m n}\frac{1}{k}-\frac{m-1}{2n},
\end{align}
where $m$ and $n$ are Hilbert space dimensions of two subsystems, and we assume that $n\ge m$. By expanding $S_R$ with large $n$, we obtain
\begin{align}\label{PageEE}
    S_R=\log m+ \frac{\frac{1-m^2}{2m }}{n}+\frac{\frac{1}{12}-\frac{1}{12m^2}}{n^2}+{\cal{O}}\left({1 \over n^4}\right).
\end{align}
When $m=n  = k = e^{\textbf{S}}$, this expansion becomes
\begin{align}
\label{Pageformula}
    S_R=\log k - \frac{1}{2} + \left( \frac{1}{2} + \frac{1}{12} \right)\frac{1}{k^2} +{\cal{O}}\left({1 \over k^4}\right).
\end{align}

Let us compare with the non-planar correction of entanglement entropy in the PSSY model computed in eq.~(\ref{IOPSubleadingVNE}). In eq.~(\ref{IOPSubleadingVNE}), we have $1/12 k^2$, which appears in \eqref{Pageformula}. This gives a natural prediction that the resummation of the extra-handle-in-bulk diagrams, as shown in the right figure in Fig.~\ref{fig:IOPWormhole}, should yield $1 / 2 k^2$.

How to show $1/2k^2$ by re-summing all the extra-handle-in-bulk diagrams by an explicit calculation is an open question. The main difficulty is associated with the systematic resummation of all diagrams. Note that each diagram can be explicitly calculated at least in a canonical ensemble using the Weil-Petersson volume as shown in \cite{Saad:2019lba}.  
However, if we restrict only some subsets of diagrams and assume that the others do not contribute, one can handle the resummation. For example, one may consider the following ansatz for the Schwinger-Dyson equation 
\begin{align}\label{EqRl}
\lambda R(\lambda)=k+\sum\limits_{n=1}^{\infty}Z_n\frac{R(\lambda)^n}{ k^n Z_1^n},
\end{align}
with 
\begin{align}\label{ansatz}
R(\lambda)=R^0(\lambda)+R^1(\lambda), \;\;\; Z_n=Z^{\text{Disk}}_n(1+a), \;\;\; \frac{Z_n^{\text{Disk}}}{(Z_1^{\text{Disk}})^n}=e^{-\textbf{S}(n-1)},
\end{align}
where $R^0(\lambda)$ is the resolvent in the planar limit (\ref{Rplanar}). We introduce the subleading terms $R^1(\lambda)$ and $a$, where $a$ does not depend on $\lambda$ and it captures the effects of extra-handle-in-bulk on a disk. Then, we can solve $R^1(\lambda)$ perturbatively as a function of $a$, which depends on $E$ in the microcanonical ensemble. Since the black hole entropy $\textbf{S}$ depends on $E$, $E$ is a function of the dimension of the Hilbert space of the subsystem, and thus, $E$ dependence can be converted into $e^{\textbf{S}}$ dependence. See Appendix \ref{sec:appen} for more detail. 

\begin{figure}[t]
\center \includegraphics[width=3.5cm]{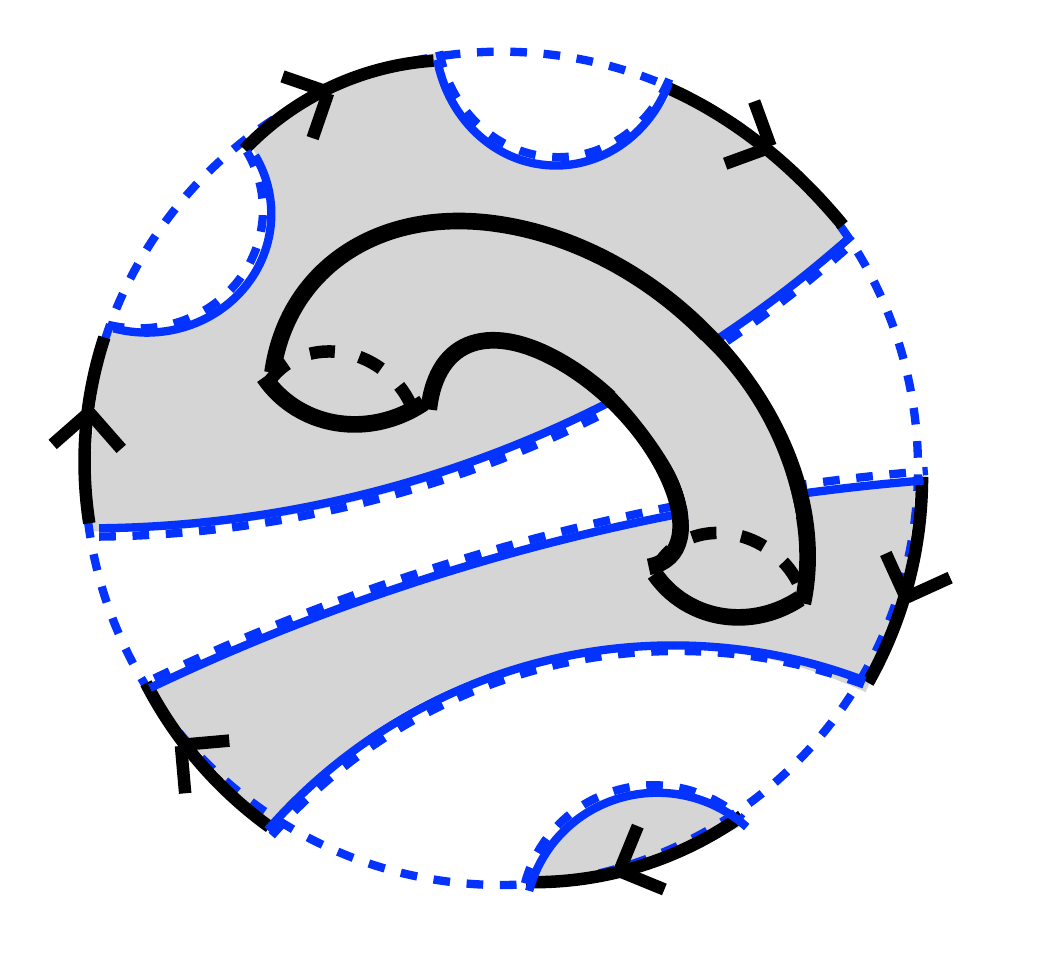}
\caption[]{An example of bulk geometries with a handle connecting two shaded regions.} 
\label{fig:TwodiskOnehandle}
\end{figure}

Of course, this approach enables us to resum only the subsets of all the diagrams with an extra handle, since the ansatz \eqref{ansatz} includes only a disk geometry with an extra handle on it.  It does not include two disks connected by a handle such as the double trumpet geometry. For example, the diagram as Fig.~\ref{fig:TwodiskOnehandle} is missing. We leave this resummation problem of all the extra-handle-in-bulk diagrams as a future problem. 

So far we have considered the correspondence in the case of $k=e^{\bold{S}}$. To generalize it to the case of $k\ne e^{\bold{S}}$ with $y = 1$, we can consider a rectangular model such that $A^\dagger$ is a rectangular $N\times K$ matrix. The two-point function $\tilde G(\omega)$ of the rectangular model in the large $N$ limit with fixed $K/N$ was derived by \cite{Iizuka:2008eb}
\begin{align}
\tilde G(\omega)=&\,\frac{i}{2\omega \lambda_{\text{'t Hooft}}}\left[h(N-Ky)+\omega(1-y)-(1-y)\sqrt{(\omega-\omega_+)(\omega-\omega_-)}\right],\\
\omega_{\pm}=&\,\frac{h}{1-y}\left(N+Ky\pm2\sqrt{NKy}\right).
\end{align}
In the infinite temperature limit (\ref{infiniteT}), we obtain
\begin{align}
\tilde G(\omega)=&\,\frac{i}{2 \omega\lambda_y} \left( \lambda_y \left(1-K/N\right) + \omega  - \sqrt{\left(\omega - \omega_+ \right)
\left( \omega - \omega_- \right)} \right),\\
\omega_{\pm}=&\,\lambda_y  \left( 1 \pm  \sqrt{K/N} \right)^2.
\end{align}
To compare with the PSSY model, let us define the following rescaled ones
\begin{align}
 \tilde{\tilde{G}}(\tilde{\omega}) &:=-i\lambda_yK\tilde G(\omega)=\frac{K}{2 \tilde{\omega}}\left(\left(K^{-1}-N^{-1}\right) + \tilde{\omega}  - \sqrt{\left(\tilde{\omega} - \tilde{\omega}_+ \right)
\left( \tilde{\omega} - \tilde{\omega}_- \right)} \right),\\
\tilde{\omega}&:=\frac{\omega}{\lambda_yK}, \;\;\; \tilde{\omega}_{\pm}:=\frac{\omega_\pm}{\lambda_yK}=\left( N^{-\frac{1}{2}} \pm  K^{-\frac{1}{2}} \right)^2.
\end{align}
Comparing them with eqs.~(\ref{Rplanar}) and (\ref{SpectralDensityWest1}),
there is a relationship between the parameters as follows
\begin{align}
\label{id4}
&\,\, \tilde{\omega} \, (\mbox{IOP}) \, \leftrightarrow \, \lambda \,(\mbox{PSSY})\,, \quad \tilde{\tilde{G}}(\tilde{\omega}) \, (\mbox{IOP}) \, \leftrightarrow \, \frac{1}{k}R(\lambda) \,(\mbox{PSSY})  \,, \\
&    N  \, (\mbox{IOP}) \, \leftrightarrow \,  k  \,(\mbox{PSSY})\,,  \qquad \,\,  K  \, (\mbox{IOP}) \, \leftrightarrow \,  e^{\bold{S}}  \,(\mbox{PSSY})\,.
\label{id5}
\end{align}

In the IOP matrix model, there is a parameter $y$ for finite temperature. However, there is no such parameter in the PSSY model, and thus we consider the infinite temperature limit (\ref{infiniteT}). It is interesting to generalize the PSSY model for the correspondence in the case of $y\ne1$.

\acknowledgments
The work of NI was supported in part by JSPS KAKENHI Grant Number 18K03619, MEXT KAKENHI Grant-in-Aid for Transformative Research Areas A “Extreme Universe” No. 21H05184. M.N. was supported by the Basic Science Research Program through the National Research Foundation of Korea (NRF) funded by the Ministry of Education (RS-2023-00245035).

\appendix

\section{Comments on the partial resum approach in the PSSY model}\label{sec:appen}

In the PSSY model, the subleading non-planar corrections of $S_\bold{R}$ come from two types of geometries such as the middle and right figures in Fig.~\ref{fig:IOPWormhole}. We estimate the non-planar correction from the geometry with an extra handle on a disk by using a simple ansatz.
Note that our ansatz does not include all geometries with an extra handle. We consider a disk geometry with an extra handle and take the partial resum of only these effects among all non-planar geometries.
We do not consider two disks connected by a handle such as the double trumpet geometry and leave its resum and evaluation as future work.

The Schwinger-Dyson equation of $R(\lambda)$ in the PSSY model is given by 
\begin{align}\label{EqRl2}
\lambda R(\lambda)=k+\sum\limits_{n=1}^{\infty}Z_n\frac{R(\lambda)^n}{ k^n Z_1^n}.
\end{align}
Then, we consider the following ansatz
\begin{align}\label{ansatz2}
R(\lambda)=R^0(\lambda)+R^1(\lambda), \;\;\; Z_n=Z^{\text{Disk}}_n(1+a), \;\;\; \frac{Z_n^{\text{Disk}}}{(Z_1^{\text{Disk}})^n}=e^{-\textbf{S}(n-1)},
\end{align}
where $R^0(\lambda)$ is the resolvent in the planar limit (\ref{Rplanar}). We introduce the subleading terms $R^1(\lambda)$ and $a$, where $a$ does not depend on $\lambda$. We set $k=e^\textbf{S}$, and $R^0(\lambda)$ becomes 
\begin{align}
R^0(\lambda)=\frac{k^2}{2}\left(1-\sqrt{\left(1-\frac{4}{k\lambda}\right)}\right) .
\end{align}

By substituting our ansatz (\ref{ansatz2}) into the Schwinger-Dyson equation (\ref{EqRl2}), we obtain the following perturbative equation of $R^1(\lambda)$
\begin{align}
\lambda R^1(\lambda)=a \frac{k R^0(\lambda)} {k^2-R^0(\lambda)}-a \frac{k^3 R^0(\lambda)} {(k^2-R^0(\lambda))^2}+R^1(\lambda)\frac{k^3} {(k^2-R^0(\lambda))^2},
\end{align}
where we leave only the first order terms proportional to $R^1(\lambda)$ or $a$.
Its solution is
\begin{align}
R^1(\lambda)=\frac{a k (R^0(\lambda))^2}{k^3-(k^2-R^0(\lambda))^2\lambda}=\frac{a k(k\lambda  -1)}{2 \lambda  }- \frac{a k^2  \sqrt{1-\frac{4}{k\lambda}} \left(k\lambda-3\right)}{2  
   (k\lambda -4)}.
\end{align}
The spectral density for $R^1(\lambda)$ is given by
\begin{align}
    D^1(\lambda)& :=- \frac{1}{\pi} \Im R^1(\lambda + i \epsilon)=\frac{ak^2  \sqrt{\frac{4}{k\lambda}-1} \left(k\lambda-3\right)}{2 \pi 
   (k\lambda -4)}\theta(\lambda)\theta\left(\frac{4}{k}-\lambda\right)-\frac{ak}{2}\delta(\lambda)\notag\\
   & =\left(\frac{ak^2}{2\pi}\frac{3}{4}\sqrt{\frac{4}{k\lambda}-1}-\frac{ak^2}{2\pi}\frac{1}{4}\frac{1}{\sqrt{\frac{4}{k\lambda}-1}}\right)\theta(\lambda)\theta\left(\frac{4}{k}-\lambda\right)-\frac{ak}{2}\delta(\lambda),
\end{align}
where the delta function term $-\frac{ak}{2}\delta(\lambda)$ comes from $-\frac{ak}{2\lambda}$ in $R^1(\lambda)$. Note that the branch points $\lambda=0, 4/k$ of $D^1(\lambda)$ are the same branch points of $D(\lambda)$ (\ref{SpectralDensityWestN}) in the planar limit. As explained in the case of the IOP matrix model, this property seems to come from the perturbative equation of $R^1(\lambda)$. One can confirm that
\begin{align}
\int d\lambda D^1(\lambda)=0, \;\;\; \int d\lambda D^1(\lambda)\lambda=0.
\end{align}
Correction of entanglement entropy by this spectral density is
\begin{align}
    S_{\bold{R}}^1:=-\int d\lambda D^1(\lambda) \lambda \log(\lambda)=\frac{a}{2}.
\end{align}

Let us specifically compute the value of $a$ in JT gravity. First, in a microcanonical ensemble, $Z_n^{\text{Disk}}(E)$ is given by \cite{Penington:2019kki}
\begin{align}\label{Zdisk}
&\qquad \quad Z_n^{\text{Disk, microcanonical}}(E) =e^{S_0}\rho_{\text{Disk}}(E)h(E,\mu)^n\Delta E,  \\
&\hspace{0cm}\rho_{\text{Disk}}(E) =  \frac{\sinh \left( 2 \pi \sqrt{2 E} \right) }{2 \pi^2}  \,,\quad 
\label{AX1} 
h(E,\mu) =2^{1-2\mu} {|\Gamma(\mu-\frac{1}{2}+i\sqrt{2E})|^2}.
\end{align}
Next, let us consider the bulk partition function with an extra handle, 
\begin{align}\label{Z1handle}
Z_n^{\text{1-handle, microcanonical}}=e^{-S_0}\rho_{\text{1-handle}}(E)h(E,\mu)^n\Delta E.
\end{align}
where $\rho_{\text{1-handle}}(E) $ can be obtained \cite{Saad:2019lba} using the explicit expression of the Weil-Petersson volume \cite{Mirzakhani:2006fta}  as 
\begin{align}
\label{rho1handle}
\rho_{\text{1-handle}}(E) &= \int_0^\infty b db V_{1,1}(b)  \rho_{\text{Trumpet}}(E,b) \,,  \\
V_{1,1}(b) &=\frac{1}{24}(b^2+4\pi^2) \,.
\end{align}
$\rho_{\text{Trumpet}}(E,b)$ is known explicitly \cite{Saad:2019pqd} as 
\begin{align}
\rho_{\text{Trumpet}}(E,b)&=\frac{\cos (b\sqrt{2E})}{\pi\sqrt{2E}}  \,.
\end{align}
However, the integral \eqref{rho1handle} for $b$ does not converge. To make it convergent, one can introduce a regulator $e^{- b \zeta}$ in the integrand of \eqref{rho1handle} as 
\begin{align}
\label{rho1handlewreg}
\rho^\zeta_{\text{1-handle}}(E) &= \int_0^\infty b db V_{1,1}(b)  \rho_{\text{Trumpet}}(E,b)  e^{- b  \zeta} 
= \frac{3 - 4 \pi^2 E}{48 \pi E^2 \sqrt{2 E}} + \order{\zeta^2}.
\end{align}
Thus, in the limit vanishing regulator $\zeta \to 0$, one obtain 
\begin{align}
\rho_{\text{1-handle}}(E) = \frac{3 - 4 \pi^2 E}{48 \pi E^2 \sqrt{2 E}} \,.
\label{AX2}
\end{align}

By combing (\ref{Zdisk}) and (\ref{Z1handle}), we obtain
\begin{align}
&Z_n^{\text{Disk, microcanonical}}+Z_n^{\text{1-handle, microcanonical}}\notag\\
&\qquad =e^{S_0}\rho_{\text{Disk}}(E)h(E,\mu)^n\Delta E\left(1+\frac{\rho_{\text{1-handle}}(E)}{e^{2S_0}\rho_{\text{Disk}}(E)}\right).
\end{align}
Therefore, in JT gravity, $a$ in our ansatz (\ref{ansatz2}) is given by
\begin{align}
\label{AX3}
a=\frac{Z_n^{\text{1-handle, microcanonical}}}{Z_n^{\text{Disk, microcanonical}}}=\frac{\rho_{\text{1-handle}}(E)}{e^{2S_0}\rho_{\text{Disk}}(E)},
\end{align}
which is a function of the fixed energy $E$ in the microcanonical ensemble. Moreover, $a$ is proportional to $\frac{1}{e^{2S_0}}$ as expected.

Let us express $a$ as a function of $\textbf{S}$. From eq.~(\ref{S02E}), we obtain $E\sim \frac{(\textbf{S}-S_0)^2}{8\pi^2}$ when $E$ is large.  Therefore using \eqref{AX1}, \eqref{AX2} and \eqref{AX3}, $a$ can be expressed under this approximation of large $E$ as 
\begin{align}
a=\frac{\rho_{\text{1-handle}}(E)}{e^{2S_0}\rho_{\text{Disk}}(E)}\sim \frac{16\pi^6(6-(\textbf{S}-S_0)^2)}{3e^{\textbf{S}+S_0}(\textbf{S}-S_0)^5}. 
\end{align}
The expression of $a$ depends not only on $\textbf{S}$, which is related to the dimension of Hilbert space of the subsystem, but also on $S_0$. This result means that, in contrast to Page's conjecture (\ref{Pageformula}), $a$ cannot be expressed by the dimension of Hilbert space only. This discrepancy with Page's conjecture might be resolved by doing the resum including all geometries with an extra handle.

\bibliography{Refs}

\bibliographystyle{JHEP}

\end{document}